\documentclass[preprint,12pt]{elsarticle}

\usepackage{graphicx}
\usepackage{float}
\usepackage{epstopdf}
\usepackage{comment}
\usepackage{algorithm}
\usepackage{algorithmic}

\usepackage{amsmath}
\usepackage{amsfonts}
\usepackage{amssymb}
\usepackage{array}
\usepackage{multirow}
\usepackage[margin=1.5cm]{geometry}
\usepackage{graphicx,subcaption,lipsum}
\usepackage{hyperref}
\usepackage{varioref}
\usepackage{cleveref}
\usepackage{caption}

\usepackage{hyperref}
\hypersetup{
     colorlinks=true,
   linkcolor=blue, 
     citecolor=blue,   
  urlcolor=blue     
}

\usepackage{xcolor}
\definecolor{royalblue}{RGB}{65, 105, 225}

\journal{Integration, the VLSI Journal}

\begin{document}

\begin{frontmatter}

\title{E2AFS: Energy-Efficient Approximate Floating Point Square Rooter for Error Tolerant Computing}

\author[inst1]{Prateek Goyal}
\author[inst1]{Jatin~Kumar~Reddy~Mothe}
\author[inst1]{Swara~Rajesh~Shelke}
\author[inst1]{Sujit Kumar Sahoo}

\affiliation[inst1]{
  organization={School of Electrical Sciences, Indian Institute of Technology Goa},
  city={Ponda},
  postcode={403401},
  state={Goa},
  country={India}
}

\begin{abstract}

Floating-point square-root computation is a power- and delay-critical operation in edge-AI, signal-processing, and embedded systems. Conventional implementations typically rely on multipliers or iterative pipelines, resulting in increased hardware complexity, switching activity, and energy consumption. This work presents E2AFS, a lightweight and fully multiplier-free floating-point square-root architecture optimized for energy-efficient computation. By reducing logic depth and minimizing switching activity, the proposed design achieves substantial improvements in hardware efficiency and performance.
FPGA implementation on an Artix-7 device demonstrates that E2AFS achieves the lowest dynamic power (7.63\,mW), the shortest critical-path delay (4.639\,ns), and the minimum power--delay product (35.39\,pJ) compared to existing ESAS and CWAHA architectures. Error evaluation using multiple accuracy metrics, together with graphical analysis, shows that E2AFS closely approximates the exact square-root function with consistently low deviation. Application-level validation in Sobel edge detection and K-means color quantization further confirms its suitability for low-power real-time edge and embedded platforms.
\end{abstract}

\end{frontmatter}

\section{Introduction}
Approximate computing has emerged as an effective design strategy for modern electronic and communication systems that must meet high-throughput demands under stringent power and resource constraints. With the rapid proliferation of edge AI, IoT sensor nodes, real-time vision modules, and embedded automation, exact computation often results in excessive energy consumption and substantial hardware overhead~\cite{AC1}. With energy efficiency becoming a central requirement from battery-powered edge devices to data-intensive platforms, approximation offers a practical and scalable path toward designing sustainable, high-performance systems for resource-constrained environments~\cite{AC3}. By selectively relaxing computational precision, approximate computing reduces logic complexity, switching activity, and critical-path delay, leading to significant savings in area and energy while maintaining accuracy levels acceptable for error-tolerant workloads such as image processing, machine learning, and signal analytics~\cite{AC3}.

Approximate arithmetic units have become essential for building energy-efficient digital systems, as operations like addition, multiplication, and square root estimation dominate the power budget in many electronic and communication applications \cite{AC4}. By relaxing full precision, these units simplify logic, reduce switching activity, and shorten carry propagation, leading to notable reductions in power consumption and hardware area. Such energy savings are especially critical for power-constrained platforms, including IoT nodes, edge-vision modules, and autonomous embedded systems. The resulting gains in throughput and latency further enable real-time performance on compact, resource-limited hardware \cite{AC5}.

Approximate arithmetic circuits, including adders \cite{AD1}\cite{AD2}, multipliers \cite{ML1}\cite{ML2}, and especially approximate square rooters \cite{LESQEC}\cite{ESAS}\cite{OLSR}\cite{CWAHA} are essential for building high-performance, energy-efficient digital systems used in modern electronics, communication, and embedded platforms. Since floating-point operations often dominate execution time and power consumption in signal processing and machine learning workloads, approximate floating-point square-root units \cite{ESAS}\cite{CWAHA} have become increasingly important. By eliminating multipliers and iterative decision stages, these architectures significantly reduce hardware area, power usage, and critical-path delay. Such advantages make them well-suited for edge-vision pipelines, autonomous sensing, real-time control, and inference engines, where square-root computations occur frequently. In this research work, we focus on multiplier-free, iteration-free floating-point square-root designs, which offer fast, energy-efficient computation for edge-intelligent systems.

\textbf{Motivation behind the proposed enhancement:}
Floating-point square root units are traditionally burdened by multipliers and deep iterative stages, leading to high resource utilization, large power dissipation, and increased delay, making them unsuitable for modern low-power and real-time systems. Existing designs either consume excessive LUTs or incur significant dynamic power and latency, limiting their applicability in energy-constrained edge platforms. To address these limitations, a lightweight, multiplier-free square-root architecture is needed to reduce logic complexity and switching activity. Eliminating multipliers and streamlining the computation flow enables lower area, delay, and Power–Delay Product (PDP), providing an energy-efficient solution for embedded and edge-intelligent systems.

\textbf{Research Contributions}: Motivated by the high hardware cost, latency, and energy demands of conventional floating-point square-root units, this work introduces a multiplier-free FP square-rooter that reduces power and critical-path delay while preserving practical accuracy. The main contributions are summarized as follows:

\begin{itemize}

\item \textbf{Energy-Efficient FP Square Rooter}: A fully multiplier-free architecture is proposed, substantially lowering logic complexity, hardware usage, and switching activity. The streamlined datapath reduces latency and power, yielding a markedly improved Power-Delay Product (PDP).

\item \textbf{FPGA Validation and Analysis}: FPGA synthesis results confirm significant reductions in resource usage, delay, and power. These gains are further illustrated through comparative graphical analysis.

\item \textbf{Accuracy–Efficiency Balance}: Figures of Merit (FoMs) demonstrate an effective trade-off between accuracy and hardware efficiency. The design sustains acceptable error levels for error-tolerant tasks such as edge detection and K-means–based color quantization while delivering energy-efficient performance.
\end{itemize}

In line with these objectives, Section II presents the proposed E2AFS architecture in detail, while Section III provides FPGA synthesis results supported by graphical analysis and two Figures of Merit. Section IV discusses the application-level evaluation of the design, and Section V concludes the paper.

\section{Proposed Design}
In this work, we present an Energy-Efficient Approximate Floating-Point Square Rooter (E2AFS) designed for the IEEE-$754$ half-precision floating-point format (FP$16$), which consists of a $1$-bit sign, $5$-bit exponent, and $10$-bit mantissa. Owing to its compact representation and low-power datapath, FP$16$ is widely adopted in edge-AI hardware. This makes nonlinear operations such as square root ideal candidates for approximate computing, enabling reductions in area, delay, and energy while maintaining acceptable accuracy.

The proposed architecture builds upon input normalization and exponent manipulation to obtain a hardware-friendly square-root formulation. For a $2n$-bit input operand $M$, the radicand is expressed in normalized form as

\[
M = 2^{r}\,(1 + Y), \qquad 0 \le Y < 1.
\]

Here, $r$ denotes the exponent and $Y$ represents the mantissa part. 
Taking the square root on both sides and applying the binomial expansion to approximate the power term yields a simplified expression. Since the higher-order terms of the expansion contribute negligibly to accuracy while substantially increasing circuit overhead, only the first two terms are retained. 
This leads to the following approximation:

\begin{equation}
\label{eq:1}
    \sqrt{M} = 2^{r/2} (1 + Y)^{1/2} \;\approx\; 2^{r/2}\left(1 + \frac{Y}{2}\right)
\end{equation}
The above equation consists of two terms : $2^{r/2}$ as the First Term and $\left(1 + \frac{Y}{2}\right)$ as the Second Term.

\subsubsection{\textbf{First-level approximation}}

Equation \eqref{eq:1} applies directly when $r$ is even, since $2^{r/2}$ can be efficiently implemented using shift operations. However, when $r$ is odd, the exponent $r/2$ becomes fractional, and the term $2^{r/2}$ can no longer be realized through shifts alone.
Therefore, we introduce the following replacement:


\[
2^{r/2} = 2^{(r-1)/2} \times 2^{1/2}
\]

We introduce an overestimation in the first term of \eqref{eq:1} by approximating 
$2^{1/2} \approx 1 + \tfrac{1}{2}$, which produces a positive overestimation error 
($\text{Error} = 1.5 - 1.4142 \approx +0.0858$). 
This overestimated value makes the square-root result positively biased. To counteract this effect, we intentionally apply an underestimation to the second term in \eqref{eq:1} as:

\[
\left(1 + \frac{Y}{2}\right) \approx \left(1 + \frac{Y}{4}\right)
\]
This substitution introduces a negative error given by
\[
e(Y) = -\frac{Y}{4}, \qquad 0 \le e(Y) \le -0.25,
\]
With the maximum magnitude of $-0.25$ occurring for $Y \in (0,1)$, the error contributions from the two terms partially cancel each other, effectively reducing the overall mean error for odd $r$. 
As a result, when $r$ is odd, the square-root expression in~\eqref{eq:1} becomes:

 \begin{equation}
 \label{eq:2}
     \sqrt{M} \approx 2^{(r-1)/2} \left(1 + \frac{1}{2}\right)\left(1 + \frac{Y}{4}\right)
 \end{equation}

\subsubsection{\textbf{Second-level approximation}}

The function $(1+Y)^{1/2}$ is nearly linear for small $Y$ but its slope decreases 
significantly as $Y$ approaches~1, making a single linear model inadequate and 
leading to large worst-case errors. To improve accuracy, the interval is split at a breakpoint $k$ that minimizes the global MED. A fine grid search ($10^{-3}$ resolution) identifies the optimal split near $Y \approx 0.51$, where the function’s curvature increases. Although $Y=0.51$ yields slightly lower error, it requires extra comparator hardware. Choosing $Y=0.5$ provides nearly identical accuracy while greatly simplifying the design, as the threshold can be detected directly from the mantissa MSB.
 Thus, $Y=0.5$ is adopted as the practical breakpoint, balancing accuracy and hardware efficiency.

For $Y < 0.5$, the function $(1+Y)^{1/2}$ remains nearly linear, allowing the base 
approximation $(1 + Y/2)$ to achieve low error without any correction. In contrast, 
for $Y > 0.5$, the decreasing slope of the true curve introduces a systematic bias: 
even $r$ cases yield consistent overestimation, while odd $r$ cases already using 
the underestimating term $(1 + Y/4)$ show dominant negative error. Sweep-based 
analysis (up to $10^{-6}$ resolution) shows that this residual error is almost 
constant in the high-$Y$ region, making fixed offsets the most efficient remedy. 
Thus, separate constant compensation terms are applied for even $r$ and odd $r$ 
when $Y > 0.5$, cancelling the bias without extra multipliers or control logic and 
preserving a compact, energy-efficient datapath.

Thus, for even $r$ and $Y > 0.5$ the square root of the radicand $M$ with a constant compensation term to counter consistent overestimation will be:
\begin{equation}
 \label{eq:3}
    \sqrt{M} = 2^{r/2}\left(1 + \frac{Y}{2} - 0.045\right)
\end{equation}

For odd $r$, a constant compensation term is needed to offset the negative error 
introduced by the underestimation. Thus, for odd $r$ and $Y > 0.5$, the square-root 
expression becomes:

\begin{equation}
 \label{eq:4}
    \sqrt{M} 
= 2^{(r-1)/2} \left(1 + \frac{1}{2}\right)\left(1 + \frac{Y + 0.3333}{4}\right)
\end{equation}

\begin{table}[ht!]
\centering
\caption{Dual-Level Approximation Strategy for Square-Root Calculation by E2AFS}
\label{tab:two_level_approx}
\renewcommand{\arraystretch}{1.7} 
\begin{tabular}{|m{0.95cm}|m{4.1cm}|m{4.95cm}|}

\hline
\multicolumn{1}{|c|}{\textbf{r}} &
\multicolumn{1}{c|}{\boldmath{$Y < 0.5$}} &
\multicolumn{1}{c|}{\boldmath{$Y \ge 0.5$}} \\ \hline

Even & 
$2^{r/2} \left( 1 +\frac{Y}{2} \right) $ &
$2^{r/2}  \left( 1 + \frac{Y}{2} -0.045 \right)$
\\ \hline

Odd &
$2^{\frac{(r-1)}{2}}\left( 1 +\frac{1}{2} \right)\left( 1 +\frac{Y}{4} \right)$ & 
$2^{\frac{(r-1)}{2}}\left( 1 +\frac{1}{2} \right)\left( 1 +\frac{Y+0.333}{4} \right)$ 
\\  \hline

\end{tabular}
\end{table}

 Table~\ref{tab:two_level_approx} presents the proposed dual-level approximation strategy for estimating the square root of an input radicand $M$. The input domain is divided into four regions based on the parity of the exponent $r$ (even/odd) and the magnitude of the mantissa component $Y$ ($Y < 0.5$ or $Y \ge 0.5$). For even values of $r$, a simple linear approximation of the form $(1 + Y/2)$ is used, with a small constant correction of $-0.045$ applied only in the higher $Y$ region to suppress overestimation errors. For odd $r$, the approximation applies a scaling factor $(1 + \tfrac{1}{2})$ to replace the $\sqrt{2}$ term with a deliberate overestimation.
While the factor $(1 + Y/4)$ provides a corresponding underestimation to offset the overestimation arising from the $\sqrt{2}$ term, it balances the overall error. Additionally, the error compensation $(Y + 0.333)$ enhances linearity for the region $Y \ge 0.5$, further improving the approximation accuracy.

\begin{figure}[ht!]
\centering
        \includegraphics[width=0.75\textwidth]{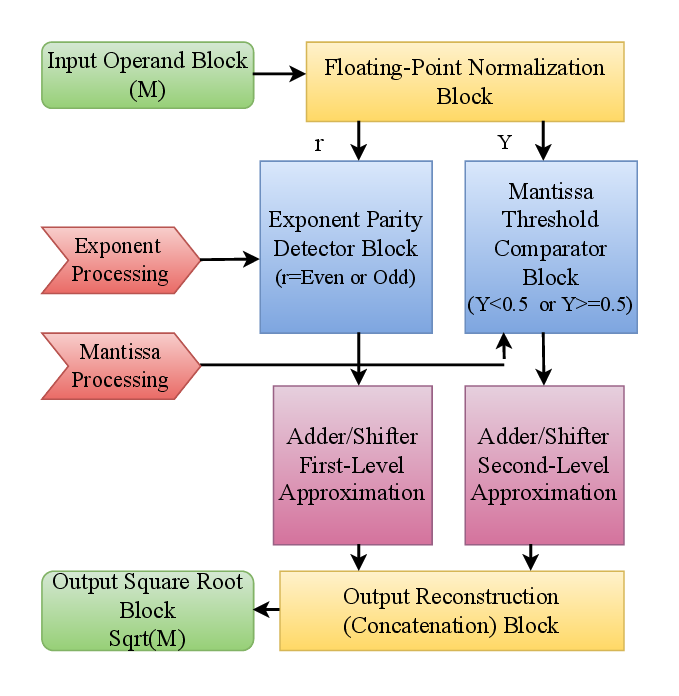}
    \caption{\small Overall architectural flow of the proposed E2AFS illustrating the key computational stages.}
    \label{fig:flow}
\end{figure}

The overall datapath of the proposed square root architecture is illustrated in Figure~\ref{fig:flow}. 
The input operand $M$ is first processed by the Floating-Point Normalization Block, which extracts the exponent–mantissa pair and generates the normalized representation required for the subsequent approximation stages. The exponent $r$ is forwarded to the Exponent Parity Detector Block, which determines whether $r$ is even or odd and selects the corresponding approximation path. In parallel, the mantissa component $Y$ is evaluated by the Mantissa Threshold Comparator Block, which identifies the appropriate region based on the condition $Y<0.5$ or $Y\ge 0.5$. These two decision blocks independently control the choice of the linearized approximation applied within the First-Level and Second-Level Adder/Shifter Approximation Units, respectively. The Output Reconstruction (Concatenation) Block assembles the final exponent and fractional components from both approximation paths to generate the final approximate square root $\sqrt{M}$. Overall, the architecture employs simple decision logic and low-cost arithmetic blocks, enabling fast and hardware-efficient computation while maintaining high numerical accuracy.

\begin{table}[ht!]
\centering
\footnotesize
\setlength{\tabcolsep}{4.5pt}
\renewcommand{\arraystretch}{1.9}

\caption{E2AFS:FP16 and Decimal Interpretation Example.}
\label{tab:example}
\begin{tabular}{|p{0.35\linewidth}|p{0.47\linewidth}|}
\hline
\multicolumn{2}{|c|}{\textbf{E2AFS: FP16 and Decimal Interpretation}} \\ \hline

\textbf{Input FP16 Representation: } &
\textbf{Input Decimal Representation:} \\[-2pt]
\(\begin{aligned}
 & \texttt{0111100001011010} \\
 M &= \texttt{0}\;
\underbrace{\texttt{11110}}_{r_1}\;
\underbrace{\texttt{0001011010}}_{y_1},
\end{aligned}\)
&
\(\begin{aligned}
35654 &=  2^{(r_1 -15)}(1+y_1) \\ &=2^{(30-15)}(1+\dfrac{90}{1024}) \\
r_1 &= 30 ,\;\; y_1 = 0.08
\end{aligned}\)\\ \hline

\textbf{Exponent Processing:} &
\textbf{Exponent Processing:} \\[-2pt]
 \(\begin{aligned}
r_1' &= \texttt{11110}-\texttt{01111} = \texttt{01111} \\
r_2'  &= \frac{r_1'-1}{2} = 00111
\end{aligned}\)
&
\(\begin{aligned}
r_1' &= 30 - 15 = 15 \\
r_2' &= \frac{15-1}{2} = 7
\end{aligned}\) \\ \hline

\textbf{Final Exponent:} &
\textbf{Final Exponent:} \\[-2pt]
\(r_2=r_2'+\texttt{01111}=\texttt{10110}\) &
\(r_2=7+15=22\) \\ \hline

\textbf{Mantissa Processing:} &
\textbf{Mantissa Processing:} \\[-2pt]
\(\begin{aligned}
r_1'\text{ odd and  }  {y_1 <0.5} \\
y_1/4 = \texttt{0000010110} 
\end{aligned}\)
&
\(\begin{aligned}
y_1 &= 0.08 < 0.5   \\
\hspace{3mm} y_1/4 &= 0.02 
\end{aligned}\) \\ \hline
\textbf{Adder/Shifter Approximation:} &
\textbf{Adder/Shifter Approximation:} \\[-2pt]

\(\begin{aligned}
Result Binary: \texttt{1000100001}
\end{aligned}\)

&
\(\begin{aligned}
(1+\frac{y_1}{4})+(1+\frac{y_1}{4})(\frac{1}{2})\\  
Approximate Sum: 1 + 0.53
\end{aligned}\)\\
\hline

\textbf{Output Reconstruction FP16:} &
\textbf{Output Reconstruction Decimal:} \\[-2pt]
\(\sqrt{M}=0\;
\underbrace{10110}_{r_2}\;
\underbrace{1000100001}_{y_2}\)
&
\(\sqrt{M}=2^{(r_2 -15)}(1+y_2)=2^{(22-15)}(1+0.53)=196.125\) \\ \hline
\end{tabular}

\end{table}

Table~\ref{tab:example} provides a detailed example illustrating how the proposed E2AFS scheme derives the square root of an input radicand $M=35654$ by processing its FP$16$ representation alongside the equivalent decimal interpretation.  

\begin{table*} [ht!]  
    \centering
    \begin{center}
      \caption{\small{Comparative Performance Analysis of 16-bit Floating-Point Square Rooters: Proposed vs. State-of-the-Art}} 
     \label{tab:comparison}
   \begin{tabular}{ |m{2.7cm}| m{0.9cm}| m{0.9cm}| m{0.9cm}| m{1.4cm} |m{0.99cm}|m{1.2cm}|m{1.2cm}|m{0.95cm}|m{01.2cm}|} 
    \hline
      Design  & \centering LUTs & \centering DP & \centering CPD  & \centering PDP  & \centering MED   & MRED  & NMED & MSE & EDmax \\
          &  & \centering \scriptsize (mW) & \centering \scriptsize (ns)  & \centering \scriptsize (pJ) & & \scriptsize ($\times 10^{-2}$)  &  \scriptsize ($\times 10^{-2}$) & &   \\
        \hline
         ESAS \cite{ESAS}   & \centering $54$ &  $7.98$ & $5.242$ & $41.8312$ & \centering $0.4625$ & \centering $1.7508$  & $0.1807$ & $2.041$ & $12.33$ \\
         \hline
        CWAHA-$4$ \cite{CWAHA}   & \centering $25$ &  $8.88$ & $5.027$ & $44.6398$ & \centering $0.5436$ & \centering $2.1823$  & $0.2124$ & $2.079$ & $11.34$ \\
         \hline
         CWAHA-$8$ \cite{CWAHA}   & \centering $45$ &  $9.99$ & $5.732$ & $57.2627$ & \centering $0.2891$ & \centering $1.1436$  & $0.1129$ & $0.899$ & $8.68$ \\
         \hline
         \textbf{E2AFS}   & \centering $37$ &  $\mathbf{7.63}$ & $\mathbf{4.639}$ & $\mathbf{35.3955}$ & \centering $0.4024$ & \centering $1.5264$  & $0.1572$ & $1.414$ & $9.98$ \\
         \hline
                           
    \end{tabular}
     \end{center}
\end{table*}

\section{Result Analysis}
All floating-point square rooter architectures were implemented in gate-level Verilog and synthesized using Xilinx Vivado~$2019.2$ on the Artix-$7$ FPGA $(XC7A35TCPG236-1)$ to evaluate hardware efficiency. The area requirement was evaluated using Look-Up Table (LUT) utilization, while dynamic power (DP) was estimated through SAIF-based post-implementation analysis using switching activity from functional simulations with varied radicand inputs. All designs were assessed under identical conditions, including the same clock frequency and supply voltage. Critical path delay (CPD) was obtained through timing analysis, and the power-delay product (PDP) was used as the primary metric for assessing energy efficiency \cite{FN2}. Accuracy was evaluated using exhaustive Verilog-based simulation in Xilinx Vivado across the complete $2^n$ input space.
 Standard error metrics, including the Mean Error Distance (MED), Mean Relative Error Distance (MRED), Normalized Mean Error Distance (NMED), Mean Squared Error (MSE), and Maximum Error Distance (EDmax), were calculated to quantify absolute, relative, and normalized error \cite{FN3}.

Table~\ref{tab:comparison} clearly demonstrates the superior performance of the proposed E2AFS architecture compared to existing non-multiplier floating-point square-rooter designs such as ESAS \cite{ESAS}, CWAHA-$4$ \cite{CWAHA}, and CWAHA-$8$ \cite{CWAHA}. The proposed design achieves the lowest dynamic power consumption ($7.63$~mW) and the shortest critical-path delay ($4.639$~ns), resulting in a significantly reduced power--delay product (PDP) of only ($35.39$~pJ). This represents a notable improvement in energy efficiency compared to all baseline designs. Although E2AFS does not have the smallest LUT count, it maintains a compact hardware footprint while providing substantially better speed--power characteristics than the ESAS and CWAHA variants. E2AFS surpasses ESAS and CWAHA-$4$ in accuracy, achieving all reduced accuracy metrics values, and maintains accuracy comparable to CWAHA-$8$ with better efficiency. 

\begin{figure}[ht!]
\centering
\hspace*{-0.45cm}
\includegraphics[scale=0.75]{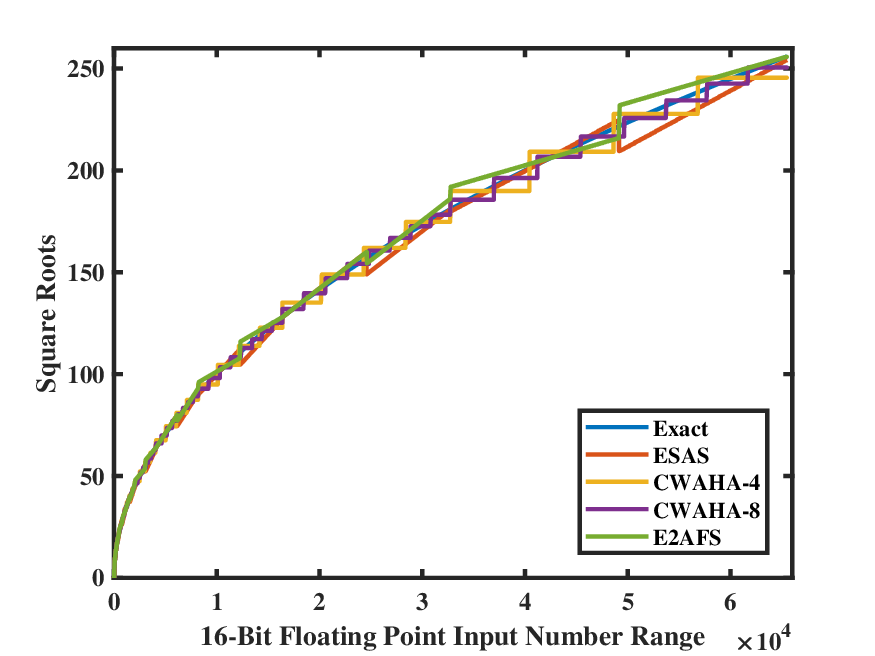}
\caption{Graphical Analysis of Various Floating-Point Square Rooters}
\label{fig:graph}
\end{figure}

The graphical plot in Figure~\ref{fig:graph} compares the square root outputs of the proposed E2AFS with ESAS, CWAHA-$4$, CWAHA-$8$, and the exact square root across the $16$-bit floating-point input range. The E2AFS curve overlaps tightly with the exact output throughout the entire range, showing minimal visible deviation. In contrast, ESAS and both CWAHA variants follow the exact trend but display noticeably larger step variations in several regions. Overall, the results indicate that all designs track the true output, but E2AFS offers the closest and most consistent approximation.

\begin{figure}[ht!]
\centering
\hspace*{-0.45cm}
\includegraphics[scale=0.75]{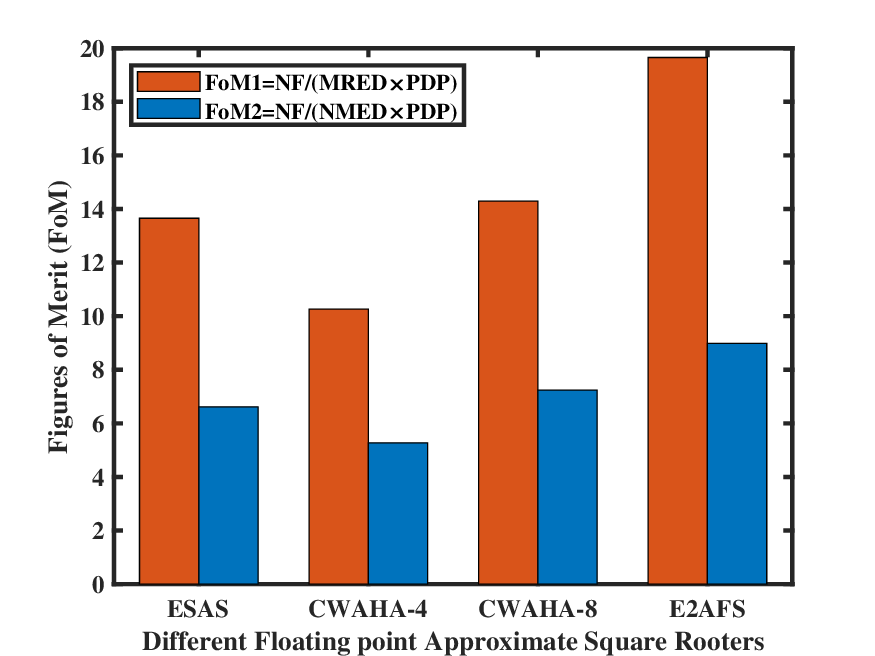}
\caption{Normalized Figures of Merit (FoM1 and FoM2) highlighting speed and energy efficiency.}
\label{fig:FoM}
\end{figure}

\begin{figure*}
	\centering
    \subfloat[\scriptsize Original]{\includegraphics[scale=0.16]{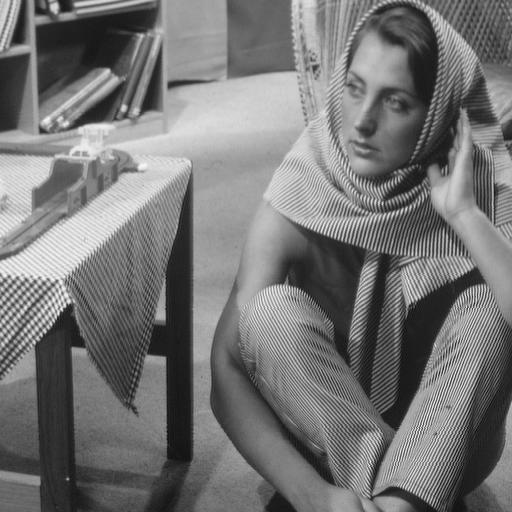}}
    \hfill	
	\subfloat[\scriptsize Exact]{\includegraphics[scale=0.16]{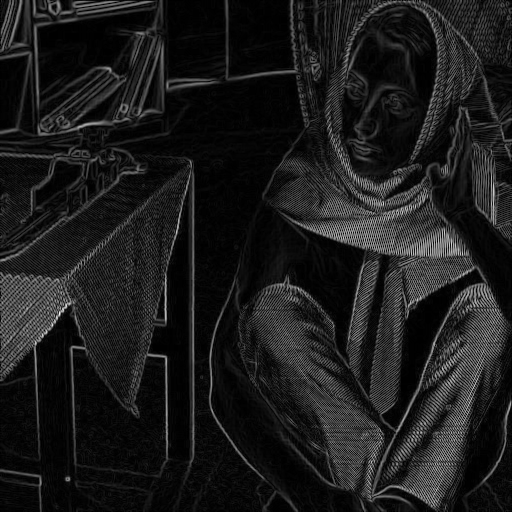}}
    \hfill
	\subfloat[\scriptsize ESAS]{\includegraphics[scale=0.16]{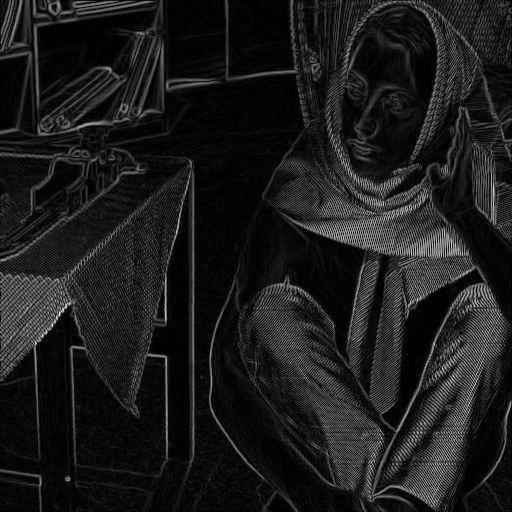}}
    \hfill
	\subfloat[\scriptsize CWAHA-4]{\includegraphics[scale=0.16]{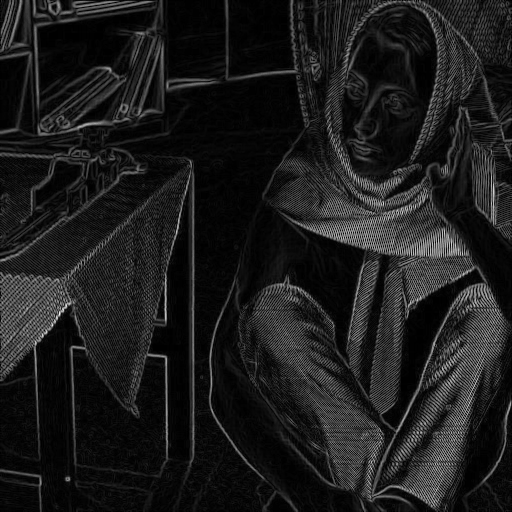}}	
      \hfill
	\subfloat[\scriptsize CWAHA-8]{\includegraphics[scale=0.16]{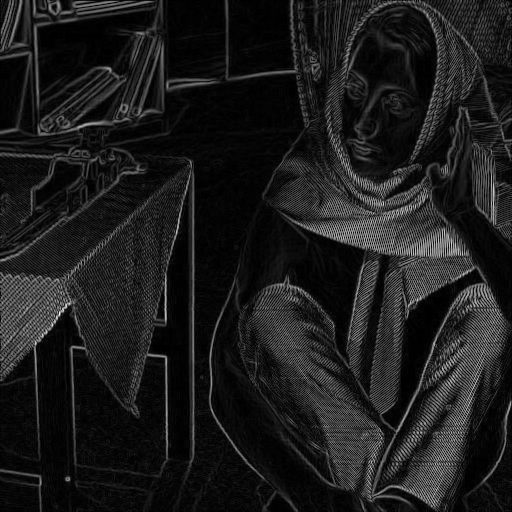}}
      \hfill
	\subfloat[\scriptsize E2AFS]{\includegraphics[scale=0.16]{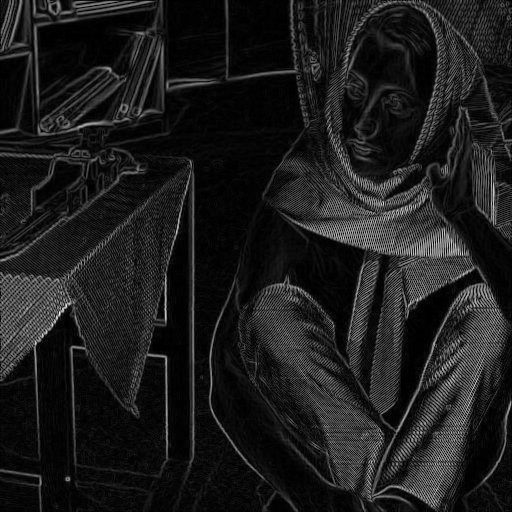}}
      \hfill
	\caption{Visual comparison of edge-detection performance across approximate square-root architectures.}
\label{fig:edge_detection_results}
\end{figure*}

Figure~\ref{fig:FoM} presents the comparative Figures of Merit (FoM$1$ and FoM$2$) for ESAS, CWAHA-$4$, CWAHA-$8$, and the proposed E2AFS design using a Normalization Factor (NF). A higher FoM signifies superior overall efficiency by jointly capturing accuracy and energy performance. As illustrated, E2AFS attains the highest FoM values in both FoM$1$ and FoM$2$  metrics, clearly surpassing all existing designs.
E2AFS achieves superior efficiency through its multiplier-free architecture and optimized datapath, reducing power, delay, and PDP while maintaining acceptable error levels. This strong balance of hardware cost, accuracy, and energy efficiency positions E2AFS as the most effective and lightweight solution among the evaluated floating-point square rooters, making it highly suitable for low-power, high-throughput, and error-tolerant edge applications.

\section{application analysis}

\subsection{Edge detection}
Due to the inherently tolerant nature of human visual interpretation, approximate computing is extensively applied in image processing and computer vision, where slight deviations in output are generally acceptable. In this experiment, the proposed square root unit is compared with an exact square root implementation within a Sobel-based edge detection framework \cite{FN4}. The Sobel operator extracts edge information by convolving the image with horizontal ($G_x$) and vertical ($G_y$) gradient masks, and the overall gradient magnitude is computed as $G = \sqrt{G_x^2 + G_y^2}$. 

\begin{table}[ht!]
\begin{center}  
\caption{\small {Assessment of Edge Detection Fidelity Using Approximate Floating-Point Square Rooters: PSNR(dB) and SSIM Metrics}}
 \label{tab:edge_comparison}
\begin{tabular}{|p{1.5cm}|p{1.6cm}|p{1.2cm}|p{2.0cm}|p{2.0cm}|p{1.2cm}|}
\hline
\centering{Quality Metrics} & \centering{ Images} & \multicolumn{4}{|p{6.4cm}|}{\centering{Approximate Floating-point Square Rooters}} \\ \hline
{} & {} & {ESAS} & {CWAHA-4} & {CWAHA-8} & {E2AFS} \\ \hline
& Peppers & $46.126$ & \centering $45.508$ & \centering $47.093$ & $46.787$  \\
& Boat & $45.012$ & \centering $44.919$ & \centering $45.929$ & $45.502$  \\
\centering {PSNR} & House & $46.783$ & \centering $46.051$ & \centering $47.936$ & $47.065$  \\
& Barbara & $45.935$ & \centering $45.019$ & \centering $46.829$ & $46.198$  \\ 

\hline
& \textbf{Average} & $45.964$ & \centering $45.374$ & \centering $46.946$ & $46.388$  \\ 
\hline
\vspace{3mm}
& Peppers & $0.9939$ & \centering $0.9913$ & \centering $0.9949$ & $0.9942$ \\
  & Boat & $0.9925$ & \centering $0.9902$ & \centering $0.9964$ & $0.9959$  \\
\centering {SSIM} & House & $0.9902$ & \centering $0.9901$ & \centering $0.9906$ & $0.9902$  \\
& Barbara & $0.9929$ & \centering $0.9909$ & \centering $0.9958$ & $0.9958$ \\

\hline
& \textbf{Average} & $0.9923$ & \centering $0.9906$ & \centering $0.9944$ & $0.9941$ 
 \\ 
\hline
\end{tabular}
\end{center}
\end{table}

To assess the perceptual impact of approximation in the Sobel pipeline, we integrated the Verilog realizations of the exact and approximate $16$-bit floating-point square rooters to compute this magnitude and evaluate Peak Signal-to-Noise Ratio and Structural Similarity Index Measure (PSNR and SSIM)  as quality metrics on four $8$-bit grayscale images (Peppers, Boat, House, and Barbara), with visual comparison for the image ``Barbara'' provided in Fig.~\ref{fig:edge_detection_results}. As summarized in Table~\ref{tab:edge_comparison},  CWAHA-$8$ achieves the highest fidelity (PSNR/SSIM), but this comes with greater hardware resources, power, and delay. In contrast, the \textbf{proposed E2AFS} delivers comparable visual quality while offering significantly better energy efficiency (lower PDP and reduced resources). 

\subsection{Color quantization}
To further demonstrate the applicability of the proposed approximate square root unit in machine learning tasks, it is evaluated within a color-quantization pipeline based on the K-Means clustering algorithm, an extensively used unsupervised method in image analysis and compression. In color quantization, K-Means groups similar RGB values into 
K=20 clusters, effectively reducing the color palette while preserving visual fidelity, which is especially valuable for storage- or bandwidth-limited systems. The algorithm iteratively partitions pixel data by computing Euclidean distances between pixels and cluster centroids, making it a suitable benchmark for assessing square-root computation efficiency \cite{kmeans}.

\begin{figure}[ht]
    \centering
    
    \begin{subfigure}[t]{0.24\textwidth}
        \centering
        \includegraphics[width=\linewidth]{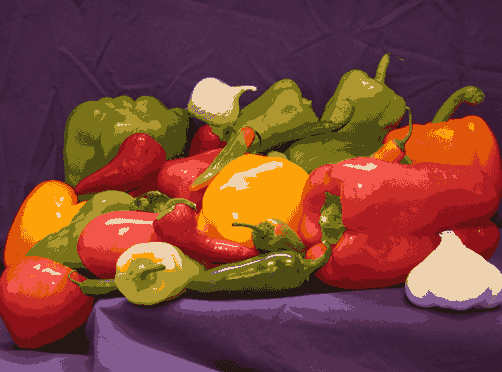}
        \caption*{\scriptsize \centering ESAS\\ PSNR=25.14dB, SSIM=0.7704}
    \end{subfigure}
    \hfill
    \begin{subfigure}[t]{0.24\textwidth}
        \centering
        \includegraphics[width=\linewidth]{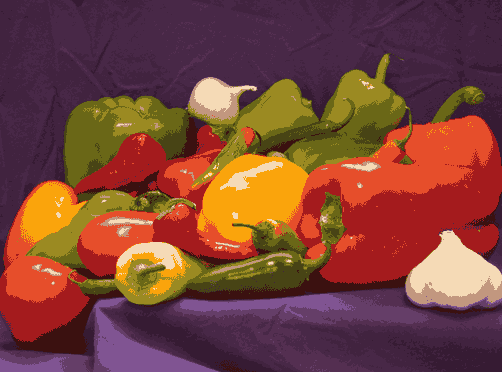}
        \caption*{\scriptsize  \centering CWAHA-4\\ PSNR=24.38dB, SSIM=0.7429}
    \end{subfigure}
    \hfill
    \begin{subfigure}[t]{0.24\textwidth}
        \centering
        \includegraphics[width=\linewidth]{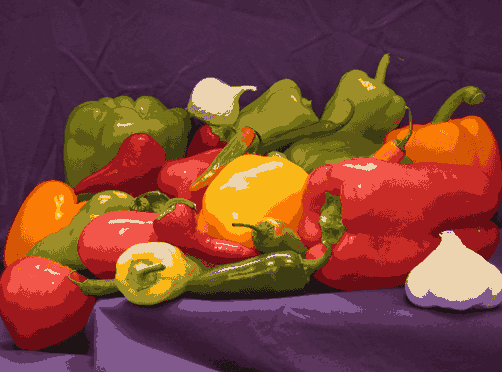}
        \caption*{\scriptsize  \centering CWAHA-8\\ PSNR=25.82dB, SSIM=0.7774}
    \end{subfigure}
    \hfill
    \begin{subfigure}[t]{0.24\textwidth}
        \centering
        \includegraphics[width=\linewidth]{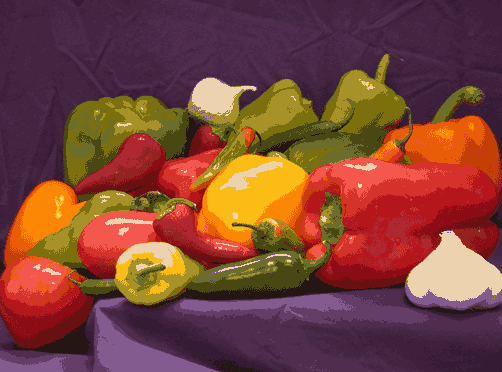}
        \caption*{\scriptsize  \centering E2AFS\\ PSNR=25.57dB, SSIM=0.7736}
    \end{subfigure}

    \caption{Visual outcomes of color quantization using various approximate square-root architectures.}
    \label{fig:kmeans_results}
\end{figure}

 For this study, the clustering process is implemented in Python, with each approximate square-root design modeled and integrated into the Euclidean distance computation of the K-Means algorithm. The proposed architecture is compared against other approximation-based square rooters previously optimized for edge-detection tasks to assess its effect on clustering accuracy and computational efficiency. As illustrated in Fig.~\ref{fig:kmeans_results}, the visual outcomes of color quantization for the ``Peppers" image show that E2AFS achieves PSNR and SSIM values that are closely aligned with the CWAHA-8 variant, while offering substantially higher energy efficiency.


\section{Conclusion}

The proposed \textbf{E2AFS} floating-point square-rooter delivers clear and compelling advantages over existing non-multiplier-based architectures in terms of hardware efficiency, computational accuracy, and energy performance. The graphical accuracy analysis demonstrates that E2AFS adheres closely to the exact square-root curve across the entire FP16 input range, exhibiting minimal deviation despite its lightweight design. As summarized in Table~\ref{tab:comparison}, E2AFS achieves the lowest dynamic power consumption, the shortest critical-path delay, and the most energy-efficient PDP among all evaluated designs, underscoring its suitability for power- and latency-sensitive microelectronic systems. These hardware gains are further supported by competitive MED, MRED, NMED, MSE, and EDmax values, confirming that the proposed approximation strategy preserves accuracy while significantly reducing complexity and switching activity. The FoM evaluation further reinforces the architecture’s superiority, with E2AFS yielding the highest FoM values and demonstrating an optimal trade-off between accuracy, delay, and power.
Collectively, these results establish E2AFS as a compact, scalable, and energy-efficient floating-point square-root unit well suited for modern VLSI and embedded systems. Its multiplier-free datapath and strong accuracy–efficiency trade-off make it ideal for edge-AI processors, IoT devices, and real-time vision platforms. With its low power, low latency, and predictable performance, E2AFS serves as an effective building block for integrated circuits and hardware accelerators where energy efficiency and area constraints are paramount.


\section*{Acknowledgment}
We thank the Visvesvaraya PhD Scheme for Electronics and IT: Phase-II (Ref.no.PhD-02/2022/25), the Science and Engineering Research Board- SERB: MTR/2021/00841, and the Indian Institute of Technology Goa (IIT Goa) for financial support. 

\bibliographystyle{IEEEtran}
\bibliography{casreference}

\end{document}